\documentclass[12pt]{article}
\usepackage{epsfig}
\begin{document}
\title{Testing the $\phi$-nuclear potential in pion-induced $\phi$ meson production on
nuclei near threshold}
\author{E. Ya. Paryev\\
{\it Institute for Nuclear Research, Russian Academy of Sciences,}\\
{\it Moscow 117312, Russia}}

\renewcommand{\today}{}
\maketitle

\begin{abstract}
The near-threshold pion-induced $\phi$ production off nuclei has been studied
in the kinematical conditions of the HADES experiment using a collision model based on the nuclear spectral function. Starting from the elementary reaction ${\pi^-}p \to {\phi}n$, absolute differential and integral cross sections for the production of $\phi$ mesons off a light carbon and a heavy tungsten target have been calculated and compared to the recently reported experimental cross sections. The absolute cross section values are shown to be sensitive to the effective nuclear scalar $\phi$ and neutron potentials while the transparency ratio is governed by the ${\phi}N$ absorption cross section. The experimental data are found to be consistent with an attractive ${\phi}$-nucleus potential of $\approx$ -(50--100) MeV at normal nuclear matter density $\rho_0$ and an effective ${\phi}N$ absorption cross section of $\approx$ 12--25 mb. The extracted $\phi$-nucleus potential is deeper than previous theoretical and experimental findings as well as those obtained in the present work within the ($t{\rho}$) approximation with accounting for Pauli
correlations from the spin-averaged  $\phi$-$p$ scattering length reported in a recent ALICE experiment.
\end{abstract}

\newpage

\section*{1 Introduction}

\hspace{0.5cm} The study of the in-medium properties (effective masses and widths) of light vector mesons $\rho$, $\omega$ and $\phi$ in nuclear matter via their production in the collisions of hadron,
heavy-ion and photon beams with nuclear targets has received considerable interest in the last few decades
(see, for example, Refs.~[1--5]) due to the hope to extract valuable information on the partial restoration
of chiral symmetry in hot or dense nuclear matter [6, 7]. This restoration is characterized by a reduction
of the scalar quark condensate $<{\bar q}q>$ in the nuclear medium compared to its vacuum value, which would
lead to a modification of hadron properties.
Among these, the $\phi$ meson attracts particular interest because of its narrow width in free space (4.3 MeV)
and because it does not overlap with other light resonances in the mass spectrum. This could allow for the
measurement of any modifications of its properties in nuclei.
Theoretically, the properties of the $\phi$ meson in nuclear matter have been extensively discussed based
on the QCD sum rules [6, 8, 9], on hadronic models [10--13] and on the Quark-Meson Coupling
model [14]. As is expected, the mass shift of the $\phi$ meson in nuclear medium at threshold is
small (about 1--3\% of its vacuum mass at normal nuclear matter density $\rho_0$), but its in-medium total
width is substantially increased (by a factor of ten at density $\rho_0$) compared to the free space value
of 4.3 MeV. The first evidence for the downward $\phi$ mass reduction of 3.4\% (of about -35 MeV)
and a width increase by a factor of 3.6 at this density was reported by the KEK-PS E325 experiment [15],
where $\phi$ mesons produced off C and Cu target nuclei with a 12 GeV proton beam have been measured.
At low energies, the production of $\phi$ mesons in nuclei by photon and proton beams has also been studied experimentally by the LEPS [16--19], CLAS [20--22] and ANKE [23] Collaborations at the SPring-8, JLab and COSY facilities, respectively. Large in-medium ${\phi}N$ absorption cross sections $\sigma_{{\phi}N}$ of about 35 mb [16],
16--70 mb [20] and 14--21 mb [23] were extracted from the comparison of the data collected in these experiments with
the respective model calculations for average $\phi$ momenta of about 1.8, 2 GeV/c and in the momentum range of
0.6--1.6 GeV/c.
These cross sections are to be compared to the total $\phi$--$N$ cross section of about 11 mb estimated by using
the vector-meson dominance model in $\phi$ photoproduction off the proton at photon energies below 10 GeV [24].
The $\phi$ production in high-energy heavy-ion collisions has been recently investigated by the STAR [25, 26],
PHENIX [27, 28] and ALICE [29] Collaborations at RHIC and LHC, respectively.
Recent reviews on the measurements of $\phi$ meson production in high-energy
heavy-ion experiments are presented in Refs. [30, 31]. The HADES Collaboration observed
deep subthreshold $\phi$ production in 1.23 A GeV Au+Au collisions [32] with a surprisingly high $\phi/K^-$
multiplicity ratio of 0.52$\pm$0.16, which indicates a possibly attractive $\phi$ in-medium mass shift.

At low energies, the production of $\phi$ mesons in nuclei has also been actively studied theoretically in
proton--nucleus [23, 33--36], photon--nucleus [24, 37--40] and heavy-ion [41] reactions, with the aim of obtaining information on a possible modification of the $\phi$ properties in cold normal and hot dense nuclear matters.
Additional information on this modification, complementary to that from proton--nucleus, photon--nucleus and
heavy-ion collisions, can be deduced from pion--nucleus reactions [5, 42]. In an attempt to get valuable
information on the behavior of $\phi$ mesons in the nuclear medium, the near-threshold
$\pi^-$ meson-induced $\phi$ production off $^{12}$C and $^{184}$W target nuclei at an incident pion momentum of
1.7 GeV/c has been recently investigated by the HADES Collaboration at SIS18/GSI [43]. For the first
time, the integrated $\phi$ production cross sections (${\Delta}{\sigma}_{\rm C}^{\phi}$
and ${\Delta}{\sigma}_{\rm W}^{\phi}$) in the HADES acceptance and the $\phi$ transparency ratio were
measured for these targets. The aim of this paper is to analyze these
data within the collision model, developed in Ref. [44] and based on the nuclear spectral function,
for incoherent direct $\phi$ production in the ${\pi^-}p \to {\phi}n$ reaction assuming different scenarios for the
${\phi}N$ absorption cross section $\sigma_{{\phi}N}$ and for the $\phi$ in-medium mass shift
(or for the $\phi$ effective scalar nuclear potential). In doing so, we will follow strictly the approach [44].
We briefly recall its main assumptions and describe, where necessary, the corresponding extensions.

\section*{2 Direct  $\phi$ meson production mechanism}

\hspace{0.5cm} The direct production of $\phi$ mesons in the kinematical conditions of the HADES experiment
in $\pi^-A$ ($A=^{12}$C and $^{184}$W) interactions at an incident pion beam momentum of 1.7 GeV/c, corresponding
to the excess energy above the ${\phi}n$ production threshold $\sqrt{s_{\rm th}}=m_{\phi}+m_n=1.959$ GeV
($m_{\phi}$ and $m_n$ are the $\phi$ meson and final neutron bare masses, respectively)
of 66 MeV, occurs in the following $\pi^-p$ elementary process
\footnote{$^)$ Our calculations showed [44] that this process dominates over secondary pion--nucleon
$\phi$ production processes in phi production off nuclei in the HADES acceptance window, corresponding to the
rapidity and transverse momentum intervals $0.4 \le y < 1.0$ and  $150 \le p_T < 650$ MeV/c, at an initial
pion momentum of 1.7 GeV/c.}$^)$:
\begin{equation}
\pi^-+p \to \phi+n.
\end{equation}
  Before going further, let us get a feeling, adopting relativistic kinematics,
about kinematic characteristics of $\phi$ mesons and neutrons allowed in this process in the
simpler case of a free target proton being at rest
at an incident pion momentum $p_{\pi^-}$ of interest. The kinematics of two-body reaction with a threshold
(as in our present case) tell us that the laboratory polar $\phi$ and final neutron
production angles $\theta_{\phi}$ and $\theta_{n}$ vary from 0 to a maximal values
$\theta^{\rm max}_{\phi}$ and $\theta^{\rm max}_{n}$, correspondingly, i.e.:
\begin{equation}
     0 \le \theta_{\phi} \le \theta^{\rm max}_{\phi},
\end{equation}
\begin{equation}
     0 \le \theta_{n} \le \theta^{\rm max}_{n};
\end{equation}
where
\begin{equation}
 \theta^{\rm max}_{\phi}={\rm arcsin}[(\sqrt{s}p^{*}_{\phi})/(m_{\phi}p_{\pi^-})],
\end{equation}
\begin{equation}
 \theta^{\rm max}_{n}={\rm arcsin}[(\sqrt{s}p^{*}_{n})/(m_{n}p_{\pi^-})].
\end{equation}
Here, the $\phi$ c.m. momentum $p^*_{\phi}$ is determined by the equation
\begin{equation}
p_{\phi}^*=\frac{1}{2\sqrt{s}}\lambda(s,m_{\phi}^{2},m_{n}^2),
\end{equation}
in which the vacuum collision energy squared $s$ and function $\lambda(x,y,z)$
are defined, respectively, by the formulas
\begin{equation}
  s=(E_{\pi^-}+m_p)^2-{p}_{\pi^-}^2,\,\,\,\,E_{\pi^-}=\sqrt{m^2_{\pi^-}+p^{2}_{\pi^-}},
\end{equation}
\begin{equation}
\lambda(x,y,z)=\sqrt{{\left[x-({\sqrt{y}}+{\sqrt{z}})^2\right]}{\left[x-
({\sqrt{y}}-{\sqrt{z}})^2\right]}},
\end{equation}
and
$p^*_{n}$ is the final neutron c.m. momentum. It is equal to the $\phi$ c.m. momentum $p^*_{\phi}$.
The quantities $m_{p}$ and $m_{\pi^-}$, entering into Eq. (7), denote the free space proton and
$\pi^-$ meson masses. From Eqs. (4), (5) one can get that
\begin{equation}
\theta^{\rm max}_{\phi}=17.427^{\circ},\,\,\,\,\theta^{\rm max}_{n}=18.963^{\circ}
\end{equation}
at an initial pion beam momentum of 1.7 GeV/c. Energy-momentum conservation in
the reaction (1), taking place in a free space, leads to two different solutions for the laboratory
$\phi$ meson and final neutron momenta $p_{\phi}$ and $p_{n}$ at given laboratory polar production angles
$\theta_{\phi}$ and $\theta_{n}$, belonging, correspondingly, to the angular intervals (2) and (3):
\begin{equation}
p^{(1,2)}_{\phi}(\theta_{\phi})=
\frac{p_{\pi^-}\sqrt{s}E^{*}_{\phi}\cos{\theta_{\phi}}\pm
(E_{\pi^-}+m_p)\sqrt{s}\sqrt{p^{*2}_{\phi}-{\gamma^2_{\rm cm}}{v^2_{\rm cm}}m^2_{\phi}\sin^2{\theta_{\phi}}}}{(E_{\pi^-}+m_p)^2-p^2_{\pi^-}\cos^2{\theta_{\phi}}},
\end{equation}
\begin{equation}
p^{(1,2)}_{n}(\theta_{n})=
\frac{p_{\pi^-}\sqrt{s}E^{*}_{n}\cos{\theta_{n}}\pm
(E_{\pi^-}+m_p)\sqrt{s}\sqrt{p^{*2}_{n}-{\gamma^2_{\rm cm}}{v^2_{\rm cm}}m^2_{n}\sin^2{\theta_{n}}}}{(E_{\pi^-}+m_p)^2-p^2_{\pi^-}\cos^2{\theta_{n}}}.
\end{equation}
Here, ${\gamma_{\rm cm}}=(E_{\pi^-}+m_p)/\sqrt{s}$, $v_{\rm cm}=p_{\pi^-}/(E_{\pi^-}+m_p)$,
$E_{\phi}^*=\sqrt{m^2_{\phi}+p^{*2}_{\phi}}$, $E^{*}_{n}=\sqrt{m^2_{n}+p^{*2}_{n}}$ and
sign "+" in the numerators of Eqs. (10), (11)
corresponds to the first solutions $p^{(1)}_{\phi}$, $p^{(1)}_{n}$
and sign "-" - to the second ones $p^{(2)}_{\phi}$, $p^{(2)}_{n}$.
Inspection of the expressions (10), (11) tells us that the first  solutions $p^{(1)}_{\phi}$ and
$p^{(1)}_{n}$ as well as the second ones $p^{(2)}_{\phi}$ and $p^{(2)}_{n}$
have different dependencies, respectively, on the production angles $\theta_{\phi}$ and $\theta_{n}$
within the angular intervals [0, $\theta_{\phi}^{\rm max}]$ and [0, $\theta_{n}^{\rm max}]$.
Namely, the former drop and the latter ones increase as the production angles
$\theta_{\phi}$ and $\theta_{n}$ increase in these intervals (cf. Fig. 1) and
\begin{equation}
p^{(1)}_{\phi}(\theta_{\phi}^{\rm max})=p^{(2)}_{\phi}(\theta_{\phi}^{\rm max})=
p_{\phi}(\theta_{\phi}^{\rm max}),
\end{equation}
\begin{equation}
p^{(1)}_{n}(\theta_{n}^{\rm max})=p^{(2)}_{n}(\theta_{n}^{\rm max})=
p_{n}(\theta_{n}^{\rm max}),
\end{equation}
where
\begin{equation}
p_{\phi}(\theta_{\phi}^{\rm max})=\sqrt{E_{\phi}^2(\theta_{\phi}^{\rm max})-m_{\phi}^2},\,\,
E_{\phi}(\theta_{\phi}^{\rm max})={\gamma_{\rm cm}}m_{\phi}^2/E_{\phi}^*;
\end{equation}
\begin{equation}
p_{n}(\theta_{n}^{\rm max})=\sqrt{E_{n}^2(\theta_{n}^{\rm max})-m_{n}^2},\,\,
E_{n}(\theta_{n}^{\rm max})={\gamma_{\rm cm}}m_{n}^2/E_{n}^*.
\end{equation}
According to Eqs. (14), (15), for $p_{\pi^-}=1.7$ GeV/c we get then that
$p_{\phi}(\theta_{\phi}^{\rm max})=0.792$ GeV/c and $p_{n}(\theta_{n}^{\rm max})=0.720$ GeV/c (cf. Fig. 1).
This figure shows that the kinematically allowed $\phi$ meson and final neutron laboratory momenta
in the process ${\pi^-}p \to {\phi}n$ on a free target proton at rest
vary within the following momentum ranges at a given initial pion momentum:
\begin{equation}
p^{(2)}_{\phi}(0^{\circ}) \le p_{\phi} \le p^{(1)}_{\phi}(0^{\circ}),
\end{equation}
\begin{equation}
p^{(2)}_{n}(0^{\circ}) \le p_{n} \le p^{(1)}_{n}(0^{\circ}),
\end{equation}
where the quantities $p^{(1,2)}_{\phi}(0^{\circ})$ and $p^{(1,2)}_{n}(0^{\circ})$
are defined, according to Eqs. (10), (11), as follows:
\begin{equation}
p^{(1,2)}_{\phi}(0^{\circ})=
{\gamma_{\rm cm}}E^{*}_{\phi}(v_{\rm cm}\pm v^{*}_{\phi}),
\end{equation}
\begin{equation}
p^{(1,2)}_{n}(0^{\circ})=
{\gamma_{\rm cm}}E^{*}_{n}(v_{\rm cm}\pm v^{*}_{n}).
\end{equation}
Here, $v^{*}_{\phi}=p^{*}_{\phi}/E^{*}_{\phi}$ and $v^{*}_{n}=p^{*}_{n}/E^{*}_{n}$.
Evidently, that
\begin{equation}
p^{(1)}_{\phi}(0^{\circ})+p^{(2)}_{n}(0^{\circ})=p_{\pi^-},\,\,\,\,
p^{(2)}_{\phi}(0^{\circ})+p^{(1)}_{n}(0^{\circ})=p_{\pi^-}.
\end{equation}
The relations (20) express the momentum conservation in the reaction (1),
taking place on a free target proton at rest, in the laboratory frame for zero $\phi$ and neutron
production angle. They correspond, respectively, to
the $\phi$ going forward (at 0$^{\circ}$) and neutron going backward (at 180$^{\circ}$)
in the c.m. system and vice versa. In line with Eqs. (18)--(19), for $p_{\pi^-}=1.7$ GeV/c we have:
\begin{equation}
p^{(1)}_{\phi}(0^{\circ})=1.217~{\rm GeV/c},\,\,\,p^{(2)}_{\phi}(0^{\circ})=0.548~{\rm GeV/c};
\end{equation}
$$
p^{(1)}_{n}(0^{\circ})=1.152~{\rm GeV/c},\,\,\, p^{(2)}_{n}(0^{\circ})=0.483~{\rm GeV/c}
$$
and inequalities (16), (17) are then
\begin{equation}
0.548~\le p_{\phi}~\le 1.217~{\rm GeV/c},
\end{equation}
\begin{equation}
0.483~\le p_{n}~\le 1.152~{\rm GeV/c}.
\end{equation}
It is obvious that the binding of target protons and their Fermi motion will distort the distributions of the
final $\phi$ mesons and neutrons as well as lead to a wider accessible
momentum intervals compared to those given above by Eqs. (22) and (23).
\begin{figure}[htb]
\begin{center}
\includegraphics[width=16.0cm]{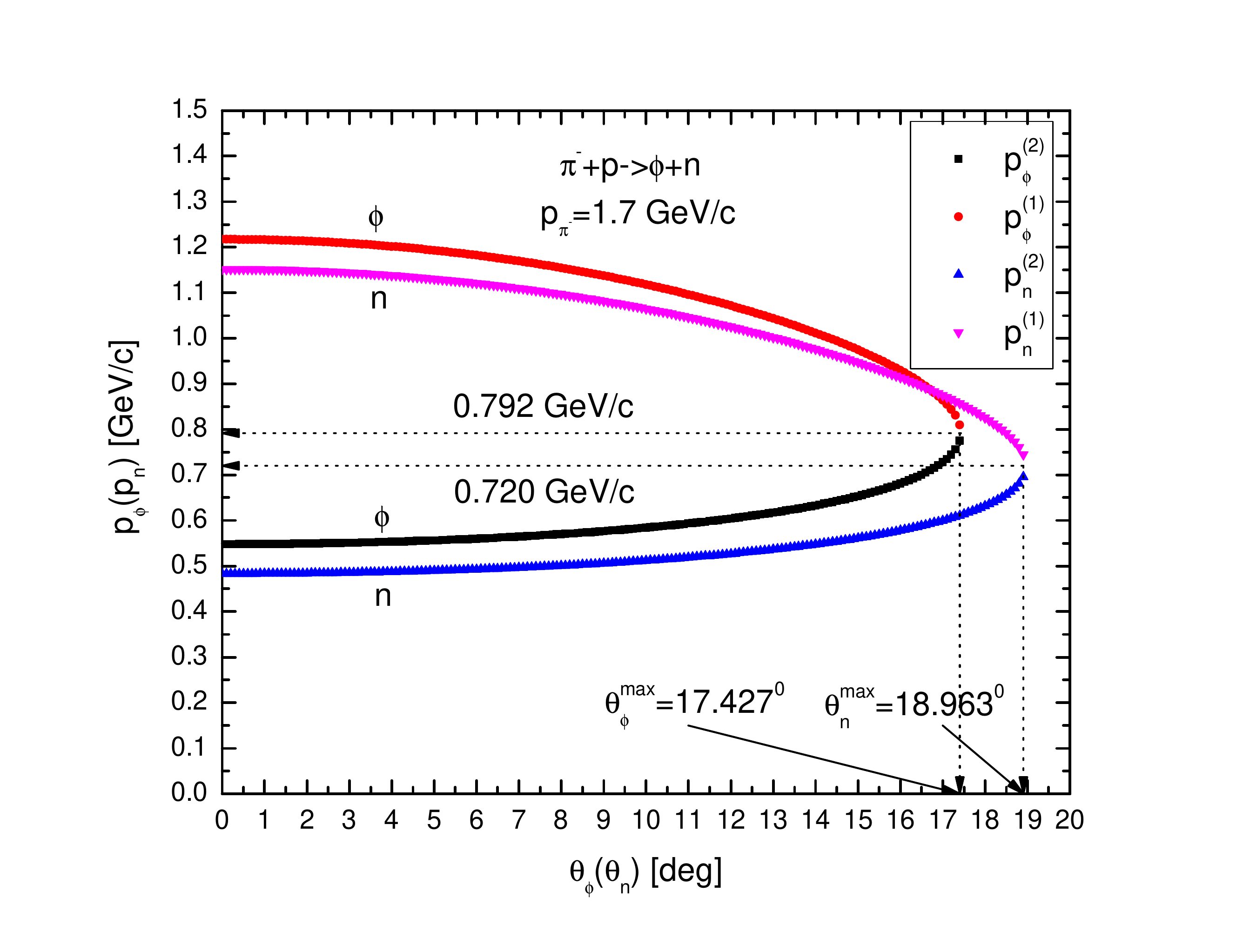}
\vspace*{-2mm} \caption{(Color online.) The kinematically allowed $\phi$ and $n$ momenta in the free space
${\pi^-}p \to {\phi}n$ reaction in the laboratory system
at an incident pion momentum of 1.7 GeV/c against their exit angles with respect to the beam direction in this system. Left and right vertical arrows mark the maximum values of these angles allowed in the reaction at given pion momentum. Upper and lower horizontal arrows indicate the values of the $\phi$ meson and  neutron momenta,
corresponding to these maximum values.}
\label{void}
\end{center}
\end{figure}

As in Ref. [44], we will approximate the in-medium local effective masses $m^*_{h}(r)$
of the final $\phi$ meson and neutron, participating in the production process (1),
with their average in-medium masses $<m^*_{h}>$ defined as:
\begin{equation}
<m^*_{h}>=m_{h}+U_h\frac{<{\rho_N}>}{{\rho_0}}.
\end{equation}
Here, $m_{h}$ ($h=\phi, n$) is the rest mass of a hadron in free space, $U_h$ is the effective scalar
nuclear potential, which the hadron sees inside the nucleus, (or its in-medium mass shift)
at normal nuclear matter density ${\rho_0}$, and $<{\rho_N}>$ is the average nucleon density.
For the target nuclei $^{12}$C and $^{184}$W,  the ratio $<{\rho_N}>/{\rho_0}$
is equal to 0.55 and 0.76, respectively.
The total energy $E^\prime_{h}$ of the hadron in nuclear matter is
expressed via its average effective mass $<m^*_{h}>$ and its in-medium momentum
${\bf p}^{\prime}_{h}$ by the equation:
\begin{equation}
E^\prime_{h}=\sqrt{({\bf p}^{\prime}_{h})^2+(<m^*_{h}>)^2}.
\end{equation}
The momentum ${\bf p}^{\prime}_{h}$ is related to the vacuum hadron momentum ${\bf p}_{h}$
as follows:
\begin{equation}
E^\prime_{h}=\sqrt{({\bf p}^{\prime}_{h})^2+(<m^*_{h}>)^2}=
\sqrt{{\bf p}^2_{h}+m^2_{h}}=E_h,
\end{equation}
where $E_h$ is the hadron total energy in vacuum.

In view of the substantial uncertainties of the $\phi$ meson self-energy at finite momenta in
the momentum range (22) (cf. Refs.[13, 45] and [46]), we will use in this work, as before in [44],
for the effective potential $U_{\phi}$ at these momenta the value corresponding to a mass drop of about -2\%
at saturation density $\rho_0$, namely $U_{\phi}=-20$ MeV, predicted at threshold by Hatsuda and Lee [6].
In addition, to extend the range of applicability of our model and to see the sensitivity of the $\phi$
production cross section of the direct process (1) to the potential $U_{\phi}$, we will yet employ in our
calculations three additional options for this potential, namely:
\begin{equation}
i)~U_{\phi}=-100~{\rm MeV},\,\,\,\,ii)~U_{\phi}=-70~{\rm MeV},\,\,\,\,iii)~U_{\phi}=-50~{\rm MeV},
\end{equation}
motivated by the following low-energy information presently available in this field.
Recently, the spin-averaged scattering length of the $p$-$\phi$ interaction was extracted
from two-particle correlations of $p$-$\phi$ pairs measured in high-multiplicity $pp$ collisions at $\sqrt{s}=13$ TeV
by the ALICE Collaboration [47]. Its real part $f_{{\phi}p}$ is found to be $f_{{\phi}p}=0.85\pm0.37$ fm
\footnote{$^)$ It is interesting to note that this (real) scattering length is an order of magnitude larger than
that obtained from the $\phi$ meson photoproduction data [22] from the CLAS experiment in Hall B of JLab for the
${\gamma}p \to {\phi}p$ reaction near threshold in Ref. [48] using the vector meson dominance, but it is comparable
within errors to that extracted in the spin 3/2 channel in Ref. [49]
adopting the lattice HAL QCD method.}$^)$.
An attractive optical low-energy $\phi$--nucleus potential depth at saturation density $\rho_0$ can be estimated,
using the impulse ($t{\rho}$) approximation [4, 50] in the large $A$ limit and accounting for the Pauli correlations term $C_{Pauli}(\rho_0)$ at this density discussed for $\Lambda$- and $\Xi^-$-nuclear potential depths, respectively, in Refs. [51] and [52, 53], as follows:
\begin{equation}
U_{\phi}=-\frac{2\pi}{m_{\phi}}\left(1+\frac{m_{\phi}}{m_N}\right)C_{Pauli}(\rho_0)f_{{\phi}p}\rho_0,
\end{equation}
where
\begin{equation}
C_{Pauli}(\rho_0)=\left[1+{\alpha_P}\frac{3k_F}{2\pi}\left(1+\frac{m_{\phi}}{m_N}\right)f_{{\phi}p}\right]^{-1},
\end{equation}
with Fermi momentum $k_F=(3\pi^2\rho_0/2)^{1/3}$. The parameter ${\alpha_P}$ in Eq. (29) switches off (${\alpha_P}=0$)
or on (${\alpha_P}=1$) Pauli correlations in ${\phi}N$ in-medium multiple scatterings.
Taking the above real part $f_{{\phi}p}$, one gets that
\begin{equation}
U_{\phi}
\approx\left\{
\begin{array}{lll}
	 -(70\pm30)~{\rm MeV}
	&\mbox{for ${\alpha_P}=0$}, \\
	&\\
        -32.0^{+5.5}_{-8.4}~{\rm MeV}
	&\mbox{for ${\alpha_P}=1$}.
\end{array}
\right.
\end{equation}
Here we have taken the density $\rho_0$ to be 0.16 fm$^{-3}$.
One sees that the $\phi$ potential depth $U_{\phi}$, obtained in the case of disregarding Pauli correlations
(${\alpha_P}=0$), decreases in size by factor of about 2--3 upon switching on these correlations (${\alpha_P}=1$).
And our choice for this depth covers fully the estimation (30) for two adopted values of the parameter
${\alpha_P}$ (see, also, the footnote 8) below).
The effective scalar neutron potential $U_n$, used in
Eq.~(24), can be determined from the relation [4, 54]
\begin{equation}
U_n=\frac{\sqrt{m_n^2+{p^{\prime}_n}^2}}{m_n}V_{NA}^{\rm SEP},
\end{equation}
where $V_{NA}^{\rm SEP}$ is the Schr${\ddot{\rm o}}$dinger equivalent potential for nucleons at normal nuclear matter density. This potential is momentum-dependent and can be parametrized as a function of the momentum relative to the
nuclear matter at rest by [54]:
\begin{equation}
V_{NA}^{\rm SEP}(p_n^{{\prime}2})=\left(V_1-V_2{\rm e}^{-2.3p_n^{{\prime}2}}\right);\,\,\,V_1=50~{\rm MeV},\,\,\,
V_2=120~{\rm MeV},
\end{equation}
where the momentum $p_n^{{\prime}}$ is measured in GeV/c. The potentials (31) and (32) are shown in Fig. 2.
\begin{figure}[htb]
\begin{center}
\includegraphics[width=16.0cm]{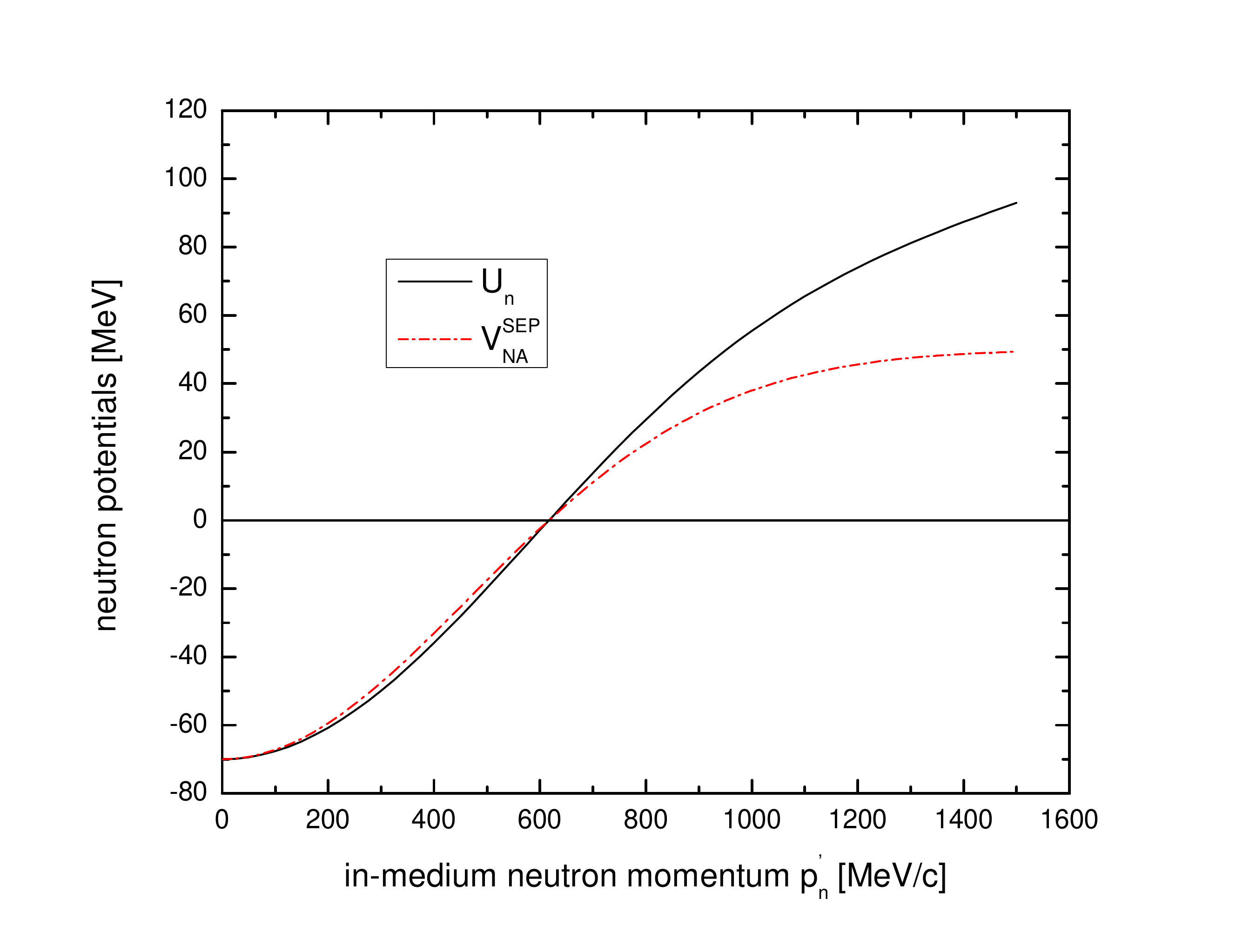}
\vspace*{-2mm} \caption{(Color online.) Momentum dependence of the effective scalar and
Schr${\ddot{\rm o}}$dinger equivalent neutron potentials at density $\rho_0$ (solid and dotted-dashed curves, respectively).}
\label{void}
\end{center}
\end{figure}
It is seen that these potentials are attractive at all momenta below of about 0.6 GeV/c with the value of
$V_{NA}^{\rm SEP}(0)=U_n(0)=-70$ MeV, whereas they are
repulsive for higher momenta and increase monotonically with increasing neutron momentum.
It is worth noting that the inclusion of the final neutron nuclear momentum-dependent potential (31)
will result in a reduction (enhancement) of the $\phi$ momentum distributions from nuclei
at low (high) meson momenta compared to the case with zero or positive neutron potential at all phi momenta
(see Figs. 3 and 4 given below). This is due to the fact that, as shown above, the low-momentum (high-momentum) $\phi$ mesons are produced in reaction (1) together with the high-momentum (low-momentum) neutrons to balance the momentum of the incident pion beam. According to Fig. 2, these neutrons feel a repulsive (attractive) potential
in the interior of the nucleus, which leads to the above effects. In Ref. [44], for the effective final neutron scalar potential $U_n$ we have used the momentum-independent positive potential with value of $U_n \approx+25$ MeV,
corresponding to its in-medium momentum $p_n^{\prime}$ of 0.78 GeV/c, determined from Eqs. (26), (31) and (32)
for the average vacuum momentum $p_n$ of 0.8 GeV/c (cf. Eq. (23)) in the kinematics of the HADES experiment.
To simplify numerical calculations, we will adopt again this potential as a basic case.
In order to see the sensitivity of the $\phi$ production differential and integral cross sections from the one-step process (1) inside the HADES acceptance to the neutron effective scalar potential $U_n$ as well as for the consistency check of the HADES data analysis, we will also use in the calculations instead the momentum-dependent one
of Eqs. (31) and (32).

In what follows for the differential cross section for $\phi$ meson production in ${\pi^-}A$ collisions
from the direct process (1) we will use Eq. (21) from Ref. [44] in these calculations, changing the limits of integration over the $\phi$ laboratory polar production angle $\theta_{\phi}$
by 0$^{\circ}$ and 90$^{\circ}$ and including in the integrand a factor $Q(p_T,y)$, which eliminates the phase
space outside of the HADES acceptance. This procedure yields for the above differential cross section the following expression:
\begin{equation}
\frac{d\sigma_{{\pi^-}A\to {\phi}X}^{({\rm prim})}
(p_{\pi^-},p_{\phi})}{dp_{\phi}}=
2{\pi}\left(\frac{Z}{A}\right)\left(\frac{p_{\phi}}{p^{\prime}_{\phi}}\right)
\int\limits_{0}^{1}d\cos{{\theta_{\phi}}}I_{V}[A,\theta_{\phi}]
\left<\frac{d\sigma_{{\pi^-}p\to {\phi}n}(p_{\pi^-},
p^{\prime}_{\phi},\theta_{\phi})}{dp^{\prime}_{\phi}d{\bf \Omega}_{\phi}}\right>_AQ(p_T,y),
\end{equation}
where
\begin{equation}
I_{V}[A,\theta_{\phi}]=A\int\limits_{0}^{R}r_{\bot}dr_{\bot}
\int\limits_{-\sqrt{R^2-r_{\bot}^2}}^{\sqrt{R^2-r_{\bot}^2}}dz
\rho(\sqrt{r_{\bot}^2+z^2})
\exp{\left[-\sigma_{{\pi^-}N}^{\rm tot}A\int\limits_{-\sqrt{R^2-r_{\bot}^2}}^{z}
\rho(\sqrt{r_{\bot}^2+x^2})dx\right]}
\end{equation}
$$
\times
\int\limits_{0}^{2\pi}d{\varphi}\exp{\left[-\sigma_{{\phi}N}A\int\limits_{0}^{l(\theta_{\phi},\varphi)}
\rho(\sqrt{x^2+2a(\theta_{\phi},\varphi)x+b+R^2})dx\right]}
$$
and the phase-space eliminating factor $Q(p_T,y)$ is defined in the following way:
\begin{equation}
Q(p_T,y)=Q(p_T){\cdot}Q(y),
\end{equation}
where
\begin{equation}
Q(p_T)
=\left\{
\begin{array}{lll}
	 1
	&\mbox{for $150 \le p_T < 650~{\rm MeV/c}$}, \\
	&\\
        0
	&\mbox{{\rm otherwise}};
\end{array}
\right.	
\end{equation}
\begin{equation}
Q(y)
=\left\{
\begin{array}{lll}
	 1
	&\mbox{for $0.4 \le y < 1.0$}, \\
	&\\
        0
	&\mbox{{\rm otherwise}}
\end{array}
\right.
\end{equation}
and
\begin{equation}
p_T=p_{\phi}\sin{\theta_{\phi}},\,\,\,\,\,y=\frac{1}{2}\ln{\frac{E_{\phi}+p_{\phi}\cos{\theta_{\phi}}}
{E_{\phi}-p_{\phi}\cos{\theta_{\phi}}}}.
\end{equation}
Here in Eq. (33),
$\left<\frac{d\sigma_{{\pi^-}p\to {\phi}n}(p_{\pi^-},
p^{\prime}_{\phi},\theta_{\phi})}{dp^{\prime}_{\phi}d{\bf \Omega}_{\phi}}\right>_A$ is the
off-shell inclusive differential cross section for the production of $\phi$ mesons and neutrons
with reduced masses $<m_{\phi}^*>$ and $<m_n^*>$ and $\phi$ mesons with in-medium momentum ${\bf p}_{{\phi}}^{'}$ in reaction (1), averaged over the Fermi motion and binding energy of the protons in the nucleus.
The quantities $\rho(r)$ and $\sigma_{{\pi^-}N}^{\rm tot}$, entering into Eq. (34), denote, respectively,
the target nucleon density, normalized to unity, and the total cross section of the free ${\pi^-}N$ interaction.
For the rest of notation see Ref. [44].
\begin{figure}[htb]
\begin{center}
\includegraphics[width=16.0cm]{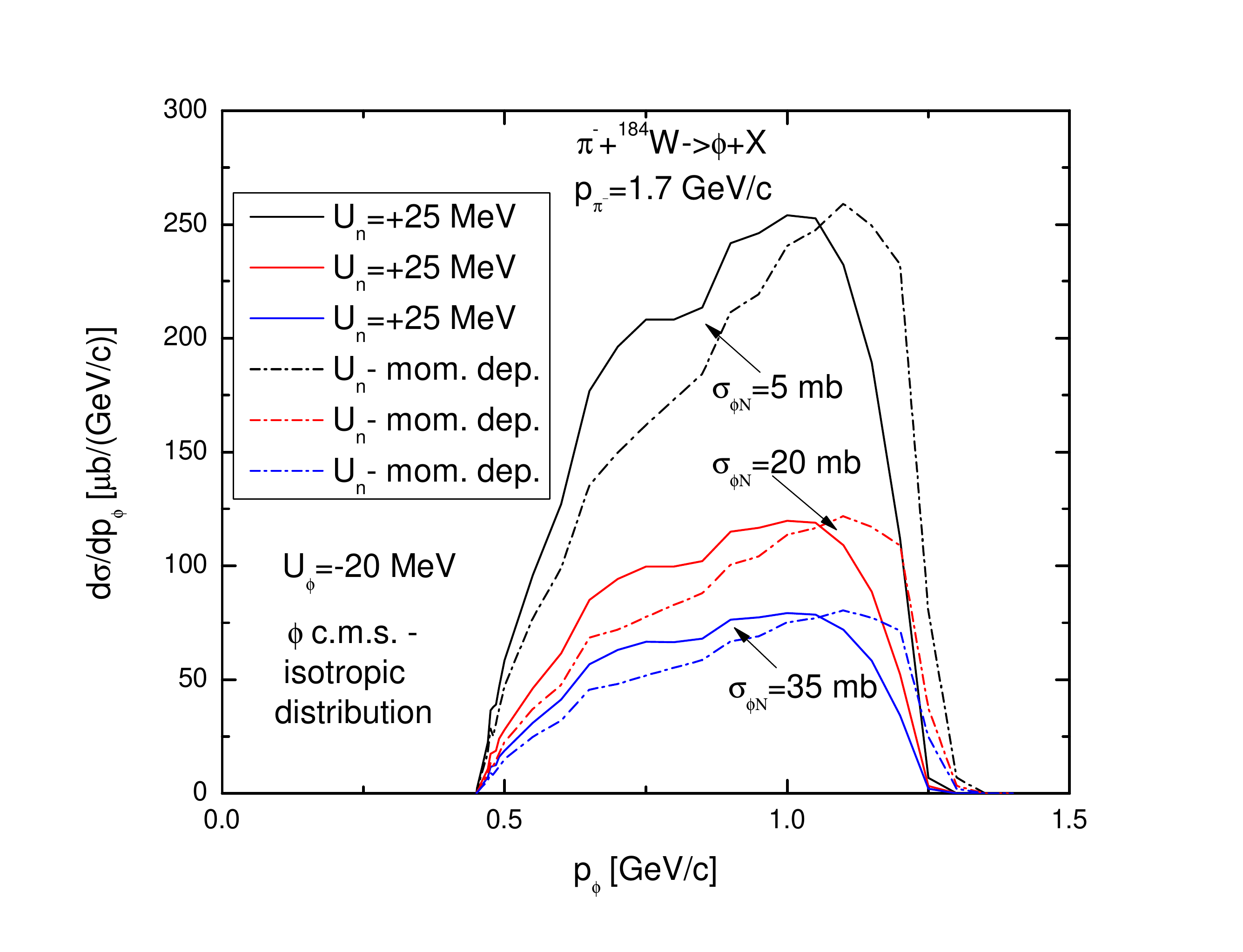}
\vspace*{-2mm} \caption{(Color online.) Momentum differential cross sections for the production of $\phi$
mesons from the primary ${\pi^-}p \to {\phi}n$ channel in the interaction of $\pi^-$ mesons with momentum
of 1.7 GeV/c with $^{184}$W nuclei for different values of the ${\phi}N$ absorption cross
section indicated in the figure, imposing (see Eqs. (36)--(38))
the HADES spectrometer kinematical cuts on the laboratory $\phi$ momenta
and production angles and for $\phi$ meson effective scalar potential $U_{\phi}=-20$ MeV at density $\rho_0$.
The solid and dashed-dotted curves are calculations, assuming that the secondary neutron feels
the momentum-independent effective scalar potential $U_{n}=+25$ MeV and the momentum-dependent potential
at this density in the nucleus (see text).}
\label{void}
\end{center}
\end{figure}
\begin{figure}[htb]
\begin{center}
\includegraphics[width=16.0cm]{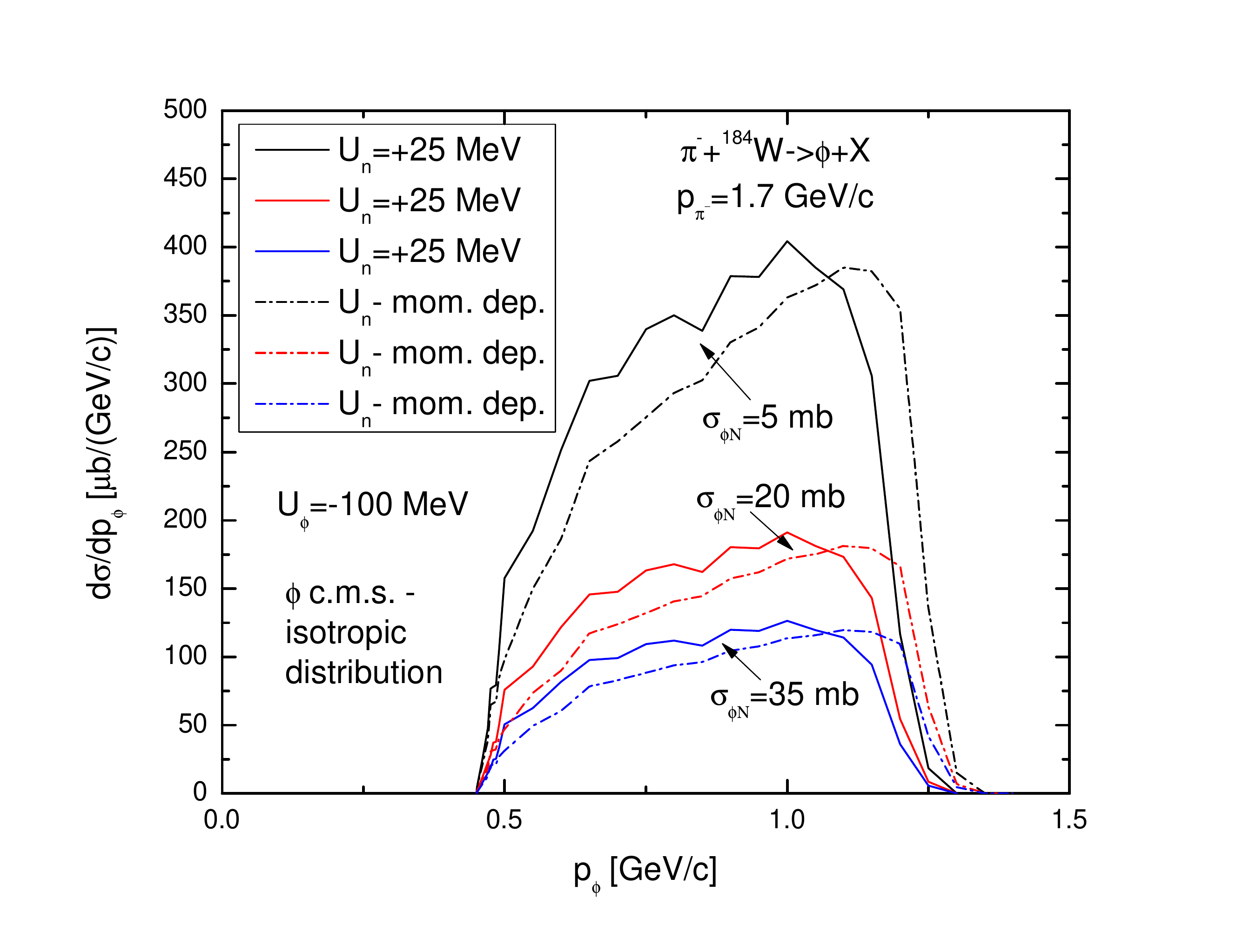}
\vspace*{-2mm} \caption{(Color online.) The same as in Fig. 3, but for the
$\phi$ meson effective scalar potential $U_{\phi}=-100$ MeV at density $\rho_0$.}
\label{void}
\end{center}
\end{figure}
\begin{figure}[htb]
\begin{center}
\includegraphics[angle=-90,width=16.0cm]{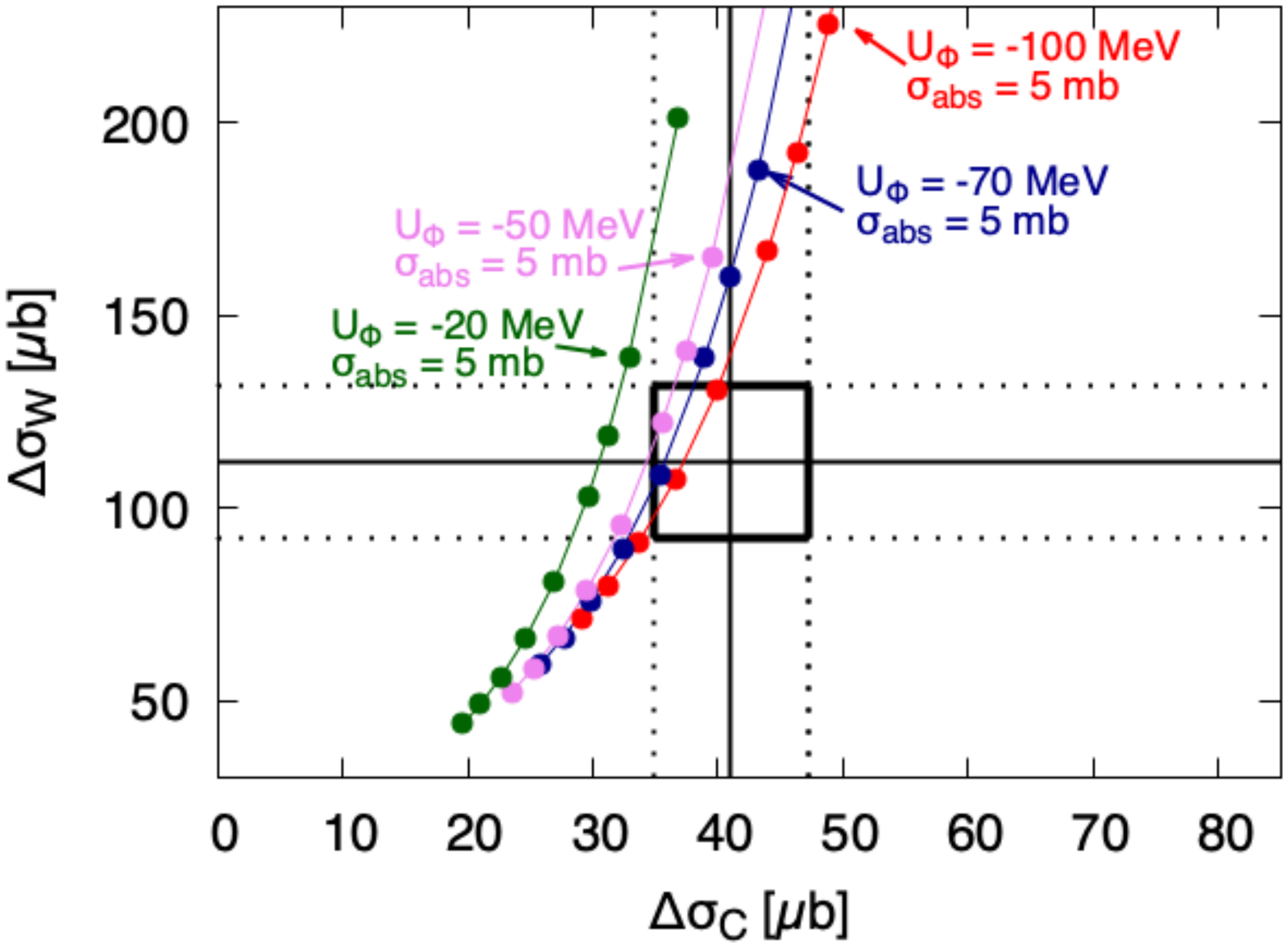}
\vspace*{-2mm} \caption{(Color online.) 2d-plot showing the calculated $\phi$ meson total production
cross sections on carbon and tungsten target nuclei in the HADES spectrometer acceptance window
for an incident $\pi^-$ meson momentum of 1.7 GeV/c for values of the ${\phi}N$ absorption cross
section of 0 (only for $U_{\phi}=-20$ MeV case), 5, 7.5, 10, 15, 20, 25, 30, 35 mb as well as for
different values of the $\phi$ meson effective scalar potential at density $\rho_0$
indicated in the plot and for secondary neutron nuclear potential $U_n=+25$ MeV at this density
(full points) in comparison with the data from the HADES experiment [43]: ${\Delta}{\sigma}_{\rm C}^{\phi}=(41\pm6.2)$
$\mu$b, ${\Delta}{\sigma}_{\rm W}^{\phi}=(112\pm19.5)$ $\mu$b. The rectangular box in the plot corresponds to these data.}
\label{void}
\end{center}
\end{figure}

In our present calculations, for the cross section  for $\phi$ meson absorption by target nucleons
$\sigma_{{\phi}N}$, entering into the expression (34), which describes (multiplied by factor $Z/A$) the effective number of protons involved in the ${\pi^-}p \to {\phi}n$ elementary reaction,
we will adopt the following representative options: $\sigma_{{\phi}N}=$ 0, 5, 7.5, 10, 15, 20, 25, 30 and 35 mb, covering in view of the aforementioned the bulk of the low-energy theoretical and experimental information presently
available in this field. The absorption cross section $\sigma_{{\phi}N}$
can be extracted in our case, in particular, from a comparison of the
total (integral) cross sections for $\phi$ meson production, calculated on the basis of Eq. (33)
inside the HADES acceptance, and measured ones [43] on $^{12}$C and $^{184}$W target nuclei
at pion beam momentum of 1.7 GeV/c, namely: ${\Delta}{\sigma}_{\rm C}^{\phi}=(41\pm6.2)$ $\mu$b and ${\Delta}{\sigma}_{\rm W}^{\phi}=(112\pm19.5)$ $\mu$b.
An alternative way to estimate this cross section
would be through a direct fit of the relative transparency ratio, measured by HADES Collaboration,
by the transparency ratio $T_A$ defined as the ratio between the total $\phi$ production
cross section in the HADES acceptance on a heavy nucleus ($^{184}$W) and on a light one ($^{12}$C), viz.:
\begin{equation}
T_A=\frac{12}{184}\frac{{\Delta}{\sigma}_{\rm W}^{\phi}}{{\Delta}{\sigma}_{\rm C}^{\phi}}.
\end{equation}
Below, we will employ these absolute and relative observables in our HADES data analysis.

\section*{3 Results and discussion}

\hspace{0.5cm} First, we consider the momentum dependencies of the absolute differential cross sections for $\phi$ meson  production from the direct process (1) in ${\pi^-}^{184}$W reactions
in the HADES acceptance window for the incident pion momentum of 1.7 GeV/c. These cross sections have been
calculated in the scenarios of collisional broadening of the $\phi$ meson characterized by the values of
$\sigma_{{\phi}N}=5$, 20 and 35 mb, for the momentum-independent positive final neutron optical potential with central value of $U_n \approx+25$ MeV and a momentum-dependent scalar nuclear potential (31), (32) and assuming an
in-medium $\phi$ mass shifts of -20 and -100 MeV at saturation density $\rho_0$. In line with Ref. [44],
the calculations have been performed, assuming the angular distribution of $\phi$ mesons
in reaction (1) to be isotropic in the ${\pi^-}p$ c.m.s. and using for its vacuum total cross section the
following expression [44]:
\begin{equation}
\sigma_{{\pi}^-p \to {\phi}n}(\sqrt{s},\sqrt{s_{\rm th}})=\left\{
\begin{array}{ll}
	0.47\left(\sqrt{s}-\sqrt{s_{\rm th}}\right)~[{\rm mb}]
	&\mbox{for $\sqrt{s_{\rm th}}<\sqrt{s}< 2.05~{\rm GeV}$}, \\
	&\\
                   23.7/s^{4.4}~[{\rm mb}]
	&\mbox{for $\sqrt{s}\ge 2.05~{\rm GeV}$}.
\end{array}
\right.	
\end{equation}
Here, $\sqrt{s}$ and $\sqrt{s_{\rm th}}$ are the free center-of-mass energy and the free threshold energy, respectively.
These dependencies are shown, respectively, in Figs. 3 and 4. They show clearly that the inclusion of the
momentum-dependent scalar nuclear potential (31), (32) leads to a modification of the $\phi$ meson momentum distributions obtained for the momentum-independent one of $U_n \approx+25$ MeV.
As noted above, the inclusion results indeed in a reduction of these distributions at $\phi$
momenta below of about 1.1 GeV/c and in their enhancement for higher momenta. However, what is very important -- the areas under the solid and dotted-dashed curves in Figs. 3 and 4 at given ${\phi}N$ absorption cross section, which are the integrated differential $\phi$ production cross sections (${\Delta}{\sigma}_{\rm W}^{\phi}$) within the HADES acceptance turned out to be almost equal to each other. Our calculations show that
the differences between them are less than 1\%. This means that the use of the final neutron
momentum-independent scalar potential of $U_n \approx+25$ MeV at density $\rho_0$ in calculating the $\phi$ total (integral) production cross sections inside the HADES acceptance of our main interest is well justified
\footnote{$^)$ In addition to this point, our calculations have shown that, for example, for $\phi$-nuclear
potential with central depth of -70 MeV increase (decrease) of the effective scalar final neutron momentum-independent
potential $U_n$ at density $\rho_0$ from $U_n=+25$ MeV to $U_n=+50$ MeV (from $U_n=+25$ MeV to $U_n=0$ MeV)
leads to reduction (enhancement) of the total cross sections for $\phi$ meson production inside the HADES
acceptance on target nuclei of interest by about 20--25\%. Such notable deviations from the $\phi$ total
production cross sections obtained for the momentum-independent neutron potential with central depth of
$U_n=+25$ MeV and, correspondingly, from the cross sections determined in the reference case of use the
momentum-dependent one (31), (32) indicate that our basic choice for final neutron optical potential is optimal.}$^)$.
It should be also pointed
out that the adopting in the calculations an anisotropic form of the $\phi$ angular distribution in the ${\pi^-}p$ c.m.s. with a slope of $b_{\phi}=2.1$ GeV$^{-2}$ as in [55] leads to only insignificant corrections of the phi total production cross section in the HADES acceptance.
The differences are about 5\%, as we found. This implies that employing a $\phi$ isotropic c.m.s. distribution in calculations of the $\phi$ integral production cross sections in ${\pi^-}A$ reactions in kinematics of the HADES experiment is also very well justified and the predictions, obtained within our approach, are robust enough.
Inspection of figures 3 and 4 tells us that the obtained results for the $\phi$ momentum distributions for $^{184}$W and, hence, those for the integrated differential $\phi$ production cross sections within the HADES acceptance
also depend strongly  on the $\phi$--nucleon absorption cross section and on the $\phi$-nuclear potential.
Thus, for example, assuming an attractive $\phi$ potential
of -100 MeV at normal nuclear matter density leads to an enhancement of the $\phi$ production cross sections
for outgoing $\phi$ meson momenta of around 1.0 GeV/c (at which they are the largest)
by factor of about 1.6 as compared to those obtained for the potential $U_{\phi}=-20$ MeV.
Since in our near-threshold calculations, contrary to the cases of proton- and photon-induced $\phi$ production in
nuclei, there is no need to know the meson creation cross section off the neutron as well as in view of the
above-mentioned, we can hope to impose reasonable constraints on these quantities using the HADES data.
\begin{figure}[!h]
\begin{center}
\includegraphics[width=16.0cm]{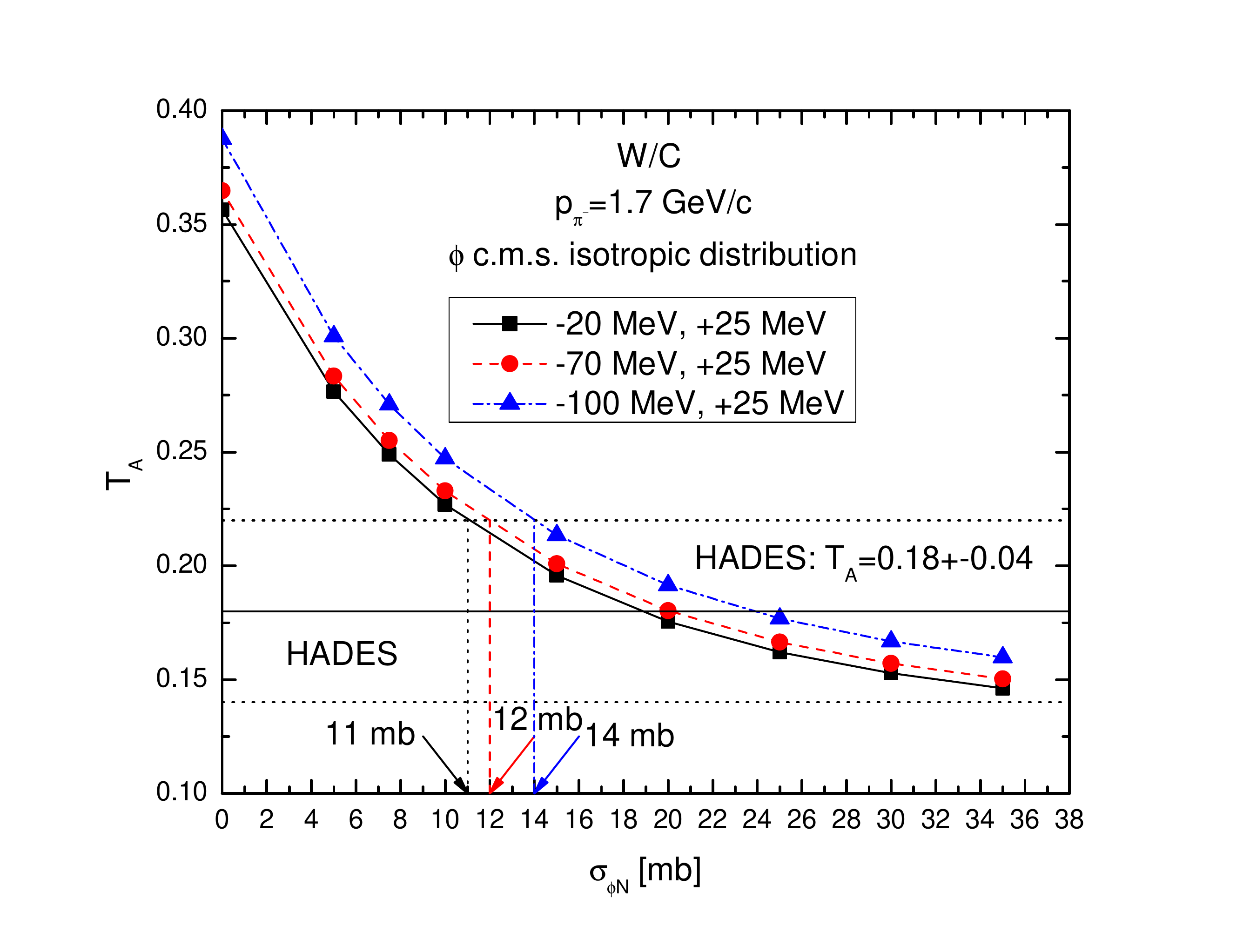}
\vspace*{-2mm} \caption{(Color online.) Transparency ratio $T_A$ as a function of the ${\phi}N$
absorption cross section for the combination $^{184}$W/$^{12}$C in the HADES spectrometer acceptance window
for an incident $\pi^-$ meson momentum of 1.7 GeV/c as well as for
different values of the $\phi$ meson effective scalar potential at density $\rho_0$
indicated in the inset and for secondary neutron nuclear potential $U_n=+25$ MeV at this density in comparison
with the value of 0.18$\pm$0.04 measured in the HADES experiment [43]. The measured central value of $T_A$ of 0.18 and error bar are indicated in the figure by horizontal solid and dotted lines, respectively.}
\label{void}
\end{center}
\end{figure}

In this context, the calculated $\phi$ meson total production cross sections on carbon and tungsten target nuclei
inside the HADES acceptance, corresponding to the $\phi$ momentum range of about 0.5--1.3 GeV/c (cf. Figs. 3, 4),
for an incident pion momentum beam of 1.7 GeV/c as well as for different values of the ${\phi}N$ absorption cross section $\sigma_{{\phi}N}$ and $\phi$-nuclear potential $U_{\phi}$ at density $\rho_0$ are shown in Fig. 5 in comparison
with the HADES data [43]. One can nicely see that the relatively shallow attractive $\phi$--nucleus potential of
$U_{\phi}$ $\simeq$ -20 MeV (or $\phi$ in-medium mass shift of about 2\%), predicted in Ref. [6] close to threshold within the QCD sum rule approach, is excluded by the data for all adopted values of the ${\phi}N$ absorption cross section $\sigma_{{\phi}N}$
\footnote{$^)$ Full points should be inside the rectangular box.}$^)$.
Looking closely at Fig. 5, one finds that the following constraints can be set for $\sigma_{{\phi}N}$ and $U_{\phi}$:
$\sigma_{{\phi}N} \ge 10$ mb and $U_{\phi} \le -50$ MeV. For realistic values of the cross section
$\sigma_{{\phi}N}$ $\approx$ 10--25 mb [21, 23, 24] a comparison of the calculated and measured $\phi$ production
cross sections indicates a sufficiently attractive $\phi$--nucleus potential with central depth of
$U_{\phi}$ $\approx$ -(50--100) MeV, which is deeper than that of $U_{\phi}$ $\approx$ -(35$\pm$7) MeV extracted
in the KEK-PS E325 experiment [15] from the data collected on a copper target in the low ${\beta}{\gamma}$ region
of $\phi$ mesons (${\beta}{\gamma} < 1.25$). Since ${\beta}{\gamma}=p_{\phi}/m_{\phi}$ and
$m_{\phi}$ $\approx$ 1.02 GeV, the condition ${\beta}{\gamma} < 1.25$ corresponds to the laboratory $\phi$ momenta
below of about 1.3 GeV/c studied also in the HADES experiment (see above). On the other hand, an attractive
$\phi$--nucleus potential of $U_{\phi}$ $\approx$ -(50--100) MeV inferred from the present analysis of the HADES
$\pi^-$--nucleus data is consistent with the potential depth of $\approx$ -(70$\pm$30) MeV estimated above
(see Eq. (30)), using the spin-averaged $\phi$-$p$ scattering length obtained from the study of the $p$-$\phi$ momentum correlations in the recent ALICE experiment [47] in the case of disregarding Pauli correlations. But it is also
deeper than that of $U_{\phi}\approx-32.0^{+5.5}_{-8.4}$ MeV obtained upon switching on these correlations.
For the proton and neutron densities, $\rho_p(r)$ and $\rho_n(r)$, in the cases of the $_{6}^{12}$C
and $_{74}^{184}$W target nuclei considered we have used in our present calculations, respectively, the harmonic
oscillator and the Woods--Saxon distributions with the same radial parameters for protons and neutrons [44].
They were normalized to $Z$ for protons ($\rho_p(r)=Z\rho(r)$) and $N=A-Z$ for neutrons ($\rho_n(r)=N\rho(r)$).
This is reasonable for light carbon nucleus, but might be questionable for heavy tungsten nucleus [51].
To see the sensitivity of the results of calculations of the $\phi$ meson total production cross sections on
the W nucleus in the HADES acceptance of our main interest to the choice of proton and neutron densities, we
adopted also for protons the two-parameter Fermi density distribution, inferred from nuclear charge distribution
for this target nucleus [56]. For the neutron density $\rho_n(r)$, which is not known to sufficient accuracy,
the 'skin' and 'halo' forms of Ref. [57] were used. Their radial parameters were determined from the r.m.s. radius
$r_n$ of $\rho_n(r)$, which assumes in line with Eq. (8)
\footnote{$^)$ In this equation the parameters $\gamma$ and $\delta$ were chosen as $\gamma=1.05$ fm,
$\delta=-0.035$ fm and the r.m.s. radius $r_p$ was set equal to that of the known nuclear charge density [56].}$^)$
of [57] larger value than that $r_p$ for proton density distribution. This replacement of the proton and neutron
densities with the same radial forms by the latter ones made little difference: the calculated $\phi$ total production
cross sections in the HADES acceptance are reduced only by a small fractions about 5 and 8\% for the 'halo' and 'skin'
shapes of $\rho_n(r)$, which cannot change our conclusion, made above, about the $\phi$-nuclear potential central
depth.

For the cross section $\sigma_{{\pi^-}N}^{\rm tot}$ entering Eq. (34) we adopt in our present calculations,
as before in Ref. [44], the value of 35 mb, representing this cross section in the case of interaction of
1.7 GeV/c pion with a free nucleon at rest [58]. But since the nucleons in the nucleus are in Fermi motion, one needs
to carry out its averaging over this motion. For this purpose, it is sufficient in principle to perform an average
of the ${\pi^-}N$ total cross section over the nucleon Fermi motion occurring along the $\pi^-$ beam axis [59]. To evaluate
the effective pion momentum range, accessible in this averaging, and the momentum dependence of the above cross
section in this range, we assume the local Fermi gas model for nucleons in the nucleus. Using for the momentum of
the target nucleon the Fermi momentum $k_F=210$ MeV/c
\footnote{$^)$ Corresponding to the average nucleon density $\rho_0/2$.}$^)$
and for its binding energy the total binding energy per nucleon $\epsilon_A=7$ MeV as well as for its total
in-medium energy the expression (11) from [44], we get that the effective initial pion momentum is 2.04 and
1.27 GeV/c for employed nucleon momentum directed, respectively, opposite and along to the incoming pion beam
\footnote{$^)$ The effective incident pion momentum is determined from the condition that the ${\pi^-}N$
center-of-mass energy squared available in the free space ${\pi^-}N$ interaction, taking place on the nucleon
at rest, is the same as that in-medium one, obtained by using Eq. (10) from Ref. [44] for the pion momentum of
1.7 GeV/c and for the considered kinematical characteristics of the intranuclear nucleon.}$^)$
.
In the pion momentum range of 1.27--2.04 GeV/c the ${\pi^-}p$ total cross section is practically
momentum-independent: here, the deviations from adopted in the calculations the value of 35 mb for it are
only about 1--5\% [58]. The ${\pi^-}n$ total cross section (which is equal to the ${\pi^+}p$ one due to the
isospin considerations) has in this momentum range a weak momentum dependence with somewhat larger but still
small deviations $\sim$ 1--15\% from the value of 35 mb also used for this cross section. It is obvious that the averaging of such weak dependencies over the Fermi motion of the intranuclear nucleons will lead only to the small corrections of the above value of 35 mb, employed in the calculations for the ${\pi^-}p$ and ${\pi^-}n$
total cross sections. These corrections are expected to be no more than the deviations indicated above. They cannot
also change our conclusion about the central depth of the $\phi$ optical potential since, for example, the change
of this value by maximal deviations $\sim$ 10--15\% leads to the alteration of the calculated $\phi$ total
production cross sections in the HADES acceptance, as the calculations have shown, by only the small fractions
of about 7--10\%.

It is interesting to consider the possibility of extracting the $\phi$--nucleon absorption cross section
$\sigma_{{\phi}N}$ from the transparency ratio defined by Eq. (39) and measured in the HADES experiment [43],
since this observable is insensitive to an in-medium massdrop of the meson [60].
Figure~6 shows the calculated transparency ratio $T_A$ for tungsten and carbon for $\phi$ mesons
produced in the primary ${\pi^-}p \to {\phi}n$ reaction channel in the HADES acceptance
by 1.7 GeV/c $\pi^-$ mesons as a function of this absorption cross section in comparison with the measured data point
of 0.18$\pm$0.04 [43]. The transparency ratio is calculated on the basis of Eq.~(39),
using the results given in Fig. 5 which were obtained for
both the adopted options for the ${\phi}N$ absorption cross section and for the $\phi$ meson and final neutron
effective scalar nuclear potentials. It is seen from this figure that the differences between all calculations corresponding to different adopted options for these potentials are indeed insignificant, especially between
options $U_{\phi}=-20$ MeV and $U_{\phi}=-70$ MeV, for all employed values of the $\phi$--nucleon absorption cross section $\sigma_{{\phi}N}$.
With this and accounting for the findings of Fig. 5, one can conclude that the measured transparency
ratio favors the ${\phi}N$ absorption cross section
$\sigma_{{\phi}N} \ge 12$ mb, the fit of its central value of 0.18 is achieved for
$\sigma_{{\phi}N}$ $\approx$ 20--25 mb for the considered $\phi$--nucleus potential central depths. The larger
values of the cross section $\sigma_{{\phi}N}$ that are required to accommodate the large experimental uncertainty
are rejected by the results given in Fig. 5 for these potential depths.

   Thus, we come to the conclusion that a consistent description of the HADES integral data presented in
Figs. 5 and 6 can be reached for realistic values of this cross section of the order of
$\sigma_{{\phi}N}$ $\approx$ 12--25 mb only if $U_{\phi}$ $\approx$ -(50--100) MeV.
The imaginary part of the $\phi$--nucleus potential $W_{\phi}$ at density $\rho_0$ can be estimated from the
expression $W_{\phi}=(1/2){\hbar}c{\beta}\sigma_{{\phi}N}{\rho_0}$,  where $\beta$ is the velocity of the $\phi$
meson in the laboratory [4]. For an average $\phi$ momentum of 800 MeV/c and
$\sigma_{{\phi}N}$ $\approx$ 20--25 mb we obtain
$W_{\phi}=$ 20--25 MeV which is a factor 2-3 smaller than the real part of the $\phi$--nucleus
potential and thus favourable for the experimental observation of $\phi$-nucleus bound
states. At least, the real part of the potential seems to be sufficiently deep to allow for the formation
and population of $\phi$ mesic states [14, 61] even if the $\phi$ meson is not produced in recoilless kinematics
\footnote{$^)$ It is worth noting that the structure and formation of the $\phi$ mesic nuclei have been
studied in Ref. [61] also for deep real $\phi$--nucleus potential with central depths of -40, -70 and -100 MeV
at threshold keeping the weak absorptive potential strength.
It was found that one can observe a clear peak structure only if this depth $\le$ -70 MeV.}$^)$.
The experimental search for such states is planned in the future J-PARC experiment E29
using antiproton annihilation on nucleus in elementary reaction ${\bar p}p \to {\phi}{\phi}$ for incident
antiproton beam momentum of 1.1 GeV/c [62, 63].
It should be also noted that the J-PARC E16 Collaboration also intends [63] to study the possible mass shift of the $\phi$ meson in cold nuclear matter in the $pA \to {\phi}X$ reaction, using a 30 GeV proton beam, more systematically via the $\phi$ dilepton decay channel with statistics two orders of magnitude higher than the KEK-PS E325 experiment [15].
Moreover, when the high intensity high resolution secondary beam line will be realized at J-PARC, an experimental study
of $\phi$ mesons in nuclear matter can be performed [63] employing the reaction ${\pi^-}A \to {\phi}nX$,
like that for $\omega$ mesons in the reaction ${\pi^-}A \to {\omega}nX$ within
the J-PARC E26 experiment.
A $\pi^-$ beam with momentum $\sim$ 2 GeV/c will be used to induce the ${\pi^-}p \to {\phi}n$ elementary reaction.
If the neutron is detected at the forward angle, slow $\phi$ mesons can be selected.
About 10 times more $\phi$ mesons compared to the E16 experiment are expected to be collected
for ${\beta}{\gamma} < 0.5$ [63].
Therefore, forward neutron measurement may lead to the observation of the $\phi$--nucleus bound states and $\phi$ in-medium modification. One may hope that the results obtained in the present work can be useful in planning these experiments.

\section*{4 Conclusions}

\hspace{0.5cm} In this paper we studied the near-threshold pion-induced production of $\phi$ mesons off nuclei
in the kinematical conditions of the HADES experiment, recently performed at GSI.
The calculations have been performed within a collision model based on the nuclear spectral function.
The model accounts for the direct ${\pi^-}p \to {\phi}n$ $\phi$ production process
as well as for the impact of the nuclear effective scalar $\phi$ and secondary neutron mean-field potentials
(or their in-medium mass shifts) on this process. We have calculated the absolute differential and integral cross sections in the HADES acceptance for the production of $\phi$ mesons off a light carbon and a heavy tungsten target
by $\pi^-$ mesons with momentum of 1.7 GeV/c. The calculations have been performed
allowing for different options for the ${\phi}N$ absorption cross section $\sigma_{{\phi}N}$
and different scenarios for the $\phi$ meson and secondary neutron effective scalar potentials at normal
nuclear matter density $\rho_0$. We also calculated the dependence of the "integral" transparency ratio for
the $\phi$ mesons on the ${\phi}N$ absorption cross section $\sigma_{{\phi}N}$ for the combination $^{184}$W/$^{12}$C.
We demonstrated that the transparency ratio for the $\phi$ mesons has, contrary to the absolute integral $\phi$
production cross sections, a weak sensitivity to the real $\phi$ meson nuclear potential. The comparison of
the calculated integral $\phi$ production cross sections and transparency ratio
with the HADES data suggests this potential to be attractive with a central value $\approx$
-(50--100) MeV for realistic values of an effective ${\phi}N$ absorption cross section of the order of $\sigma_{{\phi}N}\approx$ 12--25 mb at $\phi$ momenta of about of 0.5--1.3 GeV/c studied in the HADES experiment.
There is some tension with some theoretical and experimental findings of attractive potentials with a depth of
$\approx$ -(20--30) MeV. Further experimental and theoretical investigations on the topic of interest will be helpful
to better understand the $\phi$ meson in-medium properties.
\\
\\
{\it Note added.}-When preparing the present paper for submission to the arXiv, we became aware of
a re-analysis [64] of the experimental $p$-$\phi$ correlation function measured by ALICE [47], using
as input recent lattice calculations of the $N$-$\phi$ interaction in the spin 3/2 channel by the HAL QCD
Collaboration [49]. A constrained fit of the experimental data allows to determine the spin 1/2 channel of the
$p$-$\phi$ interaction with evidence for the formation of a  $p$-$\phi$ bound state in this channel with sufficiently
large binding energy in the range of 14.7--56.6 MeV. This supports also the findings of our present study.
\\
\\
{\bf Acknowledgments}
\\
\\
The author is very grateful to Volker Metag for initiation of this study, for detailed discussions with him during it
as well as for his essential contribution to the results presented in this publication.
\\
\\

\end{document}